\documentclass[8pt]{article}

\usepackage[a4paper, left=20mm, right=20mm, top=30mm, bottom=30mm]{geometry}
\usepackage{graphicx} 
\usepackage{caption}
\usepackage{subcaption}
\usepackage{amsmath, amsthm}
\usepackage{amsfonts}
\usepackage{import}
\usepackage{bbm}
\usepackage{algorithm}
\usepackage{algpseudocode}
\usepackage{comment}
\usepackage{siunitx}
\usepackage{pdfpages}
\usepackage{mathtools}
\usepackage{multicol}
\usepackage{bm}
\usepackage{hyperref}
\usepackage{tikz}
\usepackage{xcolor}
\usepackage{colortbl}
\usepackage{booktabs}
\usepackage[most]{tcolorbox}
\usepackage{pdfcomment}
\usepackage{dirtytalk}
\usepackage{algpseudocode}
\usepackage[utf8]{inputenc}
\usepackage{cleveref}
\usepackage{enumitem}

\newcommand{\BO}{\mathcal{O}}

\newcommand{\dt}{\Delta t} 

\newcommand{\keywords}[1]{\vspace{2mm}\noindent\textbf{Keywords:} #1}
\makeatletter
\renewcommand\@fnsymbol[1]{\@arabic{#1}}
\makeatother

\title{Numerical analysis of fluid estimation for source terms \\
in neutral particles simulation}

\author{
	Zhirui Tang\footnotemark[1] \and
	Emil Løvbak\footnotemark[2] \and
	Julian Koellermeier\textsuperscript{3,4} \and
	Giovanni Samaey\footnotemark[1]
}

\begin{document}

\date{} 

\maketitle
\vspace{-2em}

\footnotetext[1]{Department of Computer Science, KU Leuven, Leuven, Belgium. 
	Email: \texttt{zhirui.tang@kuleuven.be}}
\footnotetext[2]{Scientific Computing Center, Karlsruhe Institute of Technology (KIT), Karlsruhe, Germany}
\footnotetext[3]{Bernoulli Institute, University of Groningen, Groningen, Netherlands}
\footnotetext[4]{Department of Mathematics, Computer Science and Statistics, Ghent University, Ghent, Belgium}

\begin{abstract}
In plasma edge simulations, kinetic Monte Carlo (MC) is often used to simulate neutral particles and estimate source terms.
For large-sized reactors, like ITER and DEMO, high particle collision rates lead to a substantial computational cost for such schemes.
To address this challenge, an asymptotic-preserving kinetic-diffusion Monte Carlo (KDMC) simulation method and a corresponding fluid estimation technique have been proposed in the literature.  
In this work, we perform numerical analysis on the convergence of KDMC with the fluid estimation. 
To do so, we compare the accuracy of the analyzed algorithm with the accuracy of an approximate fluid method using the kinetic MC method as a reference.
In a one-dimensional test case, KDMC with the fluid estimation achieves at least one order of magnitude lower errors than the fluid method for both high- and low-collisional regimes.
Moreover, KDMC with the fluid estimation outperforms the kinetic MC method with a clear speed-up.
Overall, our analysis confirms the effectiveness of the discussed algorithm.
\end{abstract}


\section{Introduction}\label{sec: introduction}
We consider the modeling of neutral particles using the one-dimensional linear Boltzmann-BGK equation,
\begin{equation}\label{eqn: BBGK}
	\partial_t f(x,v,t) + v\partial_x f(x,v, t) = R(x)\left(M_p(v|x)\rho(x,v)-f(x,v,t)\right).   
\end{equation}
Here, $f(x,v,t)$ is the unknown distribution of neutral particles, $\rho(x,t)=\int f(x,v',t)dv'$ is the density with the velocity integrated over the domain $(-\infty, \infty)$ and $R(x)$ is the charge-exchange collision rate. $M_p(v|x)$ is the normalized drifting Maxwellian distribution, which is a normal distribution with mean $u_p(x)$ the mean velocity of plasma and variance $\sigma^2_p(x)=T(x)/m$, where $T(x)$ and $m$ are the temperature and the mass of plasma, respectively. 
Instead of the distribution $f(x,v,t)$, one is typically more interested in the mass, momentum, and energy sources (used to couple the neutral solver with the plasma solver in B2-EIRENE~\cite{reiterEIRENEB2EIRENECodes2005a}), 
in which the main components are three time-integrated moments, defined as
\begin{align}\label{eqn: moments}
	m_0(x) = \int_0^{\Bar{t}}\int f(x,v,t)dvdt, \ \ m_1(x) &= \int_0^{\Bar{t}} \int vf(x,v,t)dv dt, \ \ \text{and} \ \ m_2(x) = \int_0^{\Bar{t}}\int \frac{v^2}{2}f(x,v,t)dvdt.
\end{align}
The time integration up to the final simulation time $\Bar{t}$ is consistent with the time-integrated simulation strategy used in the EIRENE code~\cite{reiterEIRENEB2EIRENECodes2005a} (see the “Unbiased Estimators” section of its user manual). 
In what follows, we refer to the process of obtaining a discrete representation of the distribution $f(x,v,t)$ as simulation, while the calculation of moments \eqref{eqn: moments} is termed estimation.

The simulation of \eqref{eqn: BBGK} in EIRENE is performed using the kinetic Monte Carlo (MC) method, or simply MC method. This method is dimension-independent, facilitates handling complex geometries, and prevents the need to fully resolve the velocity domain~\cite{mortierKineticDiffusionAsymptoticPreservingMonte2022a, reiterEIRENEB2EIRENECodes2005a}. 
However, in large-sized reactors, such as ITER and DEMO, the collision rate $R(x)$ is high in certain regions of the domain~\cite{horstenComparisonFluidNeutral2016c}, leading to a high computational cost. In these regions, one can use an approximate fluid model to speed up the simulation without significant loss of accuracy~\cite{maesHilbertExpansionBased2023a}. 
Nevertheless, as the collision rate varies dramatically across the simulation domain, the low-collision regions still require the kinetic MC simulation, due to the fluid model being inaccurate in these regions.

One approach for handling the above situation is asymptotic-preserving methods. 
Our main interest lies in asymptotic-preserving Monte Carlo (APMC) methods. 
This work is based on the recently designed APMC method, known as kinetic-diffusion Monte Carlo (KDMC) \cite{mortierKineticDiffusionAsymptoticPreservingMonte2022a}. This method generates particle trajectories by combining the kinetic MC simulation with the diffusive simulation based on a fluid approximation of \eqref{eqn: BBGK}.
With the generated particle trajectories, we then estimate the moments \eqref{eqn: moments} using a fluid estimation procedure~\cite{mortierEstimationPostprocessingStep2022}, which applies 
the MC method to the kinetic part and the fluid method to the diffusive part.

In this work, we analyze the error of the resulting moments numerically.
The formulas of the final estimation error are given without proof. We plan to publish their proof in an extensive future publication. 
We refer to \cite{tang2025prep} for the details.
Our numerical analysis is performed with a one-dimensional test case used in~\cite{horstenComparisonFluidNeutral2016c}.  
Moreover, we show the effectiveness of the analyzed estimation procedure by comparing it with the fluid-based method in~\cite{maesHilbertExpansionBased2023a}.

The remainder of the paper is organized as follows. In Section \ref{sec: KDMC and fluid est}, we introduce KDMC and the associated fluid estimation procedure.
In Section \ref{sec: error analysis}, we discuss the error of the discussed method. 
Then, in Section \ref{sec: fusion test case}, we analyze the error numerically and provide the computational cost analysis. Finally, we conclude the work and give an outlook on future research in Section \ref{sec: conclusion}.

\section{KDMC, fluid estimation, and errors}\label{sec: KDMC and fluid est}
In this section, we introduce KDMC and the fluid estimation procedure. The combined algorithm is presented in Alg. \ref{alg: KDMC with f}.
Next, we give an error bound for the whole algorithm. We plan to present a full proof of this error bound in a future publication \cite{tang2025prep}.  

\subsection{KDMC simulation}\label{sec: simulation}

KDMC is parameterized by a fixed time step $\dt=\bar{t}/K$ with $K$ the number of time steps. Each $\dt$ contains a kinetic and a diffusive part. 
Consider the $i$-th particle for $i=1,2,\ldots, I$, where $I$ is the total number of particles. 
The simulation of this particle starts from the time $t=0$ at the position $x^0_i$ with the velocity $v^0_i$. 
\begin{description}[leftmargin=0pt]
	\item[Kinetic part (line \ref{line: kinetic start}-- \ref{line: kinetic end})]
	Within each $\dt$ time step, the particle first executes a free-flight with flight time $\tau$ sampled from the exponential distribution $\text{Exp}(R(x^0_i))$. The position of the particle updates to $x^1_i = x^0_i + \tau v^0_i$ (line \ref{eqn: a kinetic step}) with $\tau=\min(\tau, \dt)$, and we refer to this update as a kinetic step.
	If $\tau<\dt$, a charge-exchange collision happens and the particle is assigned a new velocity $v^1_i$ sampled from the Maxwellian distribution $M(v|x^1_i)$. 
	Otherwise, i.e., $\tau=\dt$, keep the velocity unchanged. 

	\item[Diffusive part (line \ref{line: diffusive start}--\ref{line: diffusive end})] 
	For the rest of the time $\theta=\dt-\tau$, the position is updated as $x^{1'}_i = x^1_i + A(x^1_i)\theta + \sqrt{2D(x^1_i)\theta}\xi $ (line \ref{eqn: a diffusive step}), 
	where the function $A(x)$ and $D(x)$ are derived from the fluid approximation of \eqref{eqn: BBGK}, which reads
	\begin{equation}\label{eqn: fluid-model}
		\partial_t\rho(x,t) + \partial_x \left(u_p(x)\rho(x,t)\right)-\partial_x\left(D(x)\partial_x \left(\sigma_p^2(x)\rho(x,t)\right)\right)=0,
	\end{equation}
	where $\rho(x,t)=\int f(x,v',t)dv'$ as before, and $D(x)=1/R(x)$. This positional update is called a diffusive step.
	Finally, the particle is assigned a new velocity sampled again from the distribution $M(v|x^{1'}_i)$ at the new position $x^{1'}_i$.  
\end{description}
Repeating the above process up to the time $\bar{t}$, we obtain the trajectory of the $i$-th particle, denoted as 
$\{(x_i^k, v_i^k), ({x_i^k}', {v_i^k}')\}_{k=0}^K$.
With the trajectories of all particles, we can estimate the moments \eqref{eqn: moments}.

\subsection{Estimation procedure}	
The fluid estimation procedure, corresponding to the KDMC simulation in Section \ref{sec: simulation}, also contains two parts:
\begin{description}[leftmargin=0pt]
	\item[Kinetic part (line \ref{line: mc_estimator})]
	The state $({x_i^{k}}, {v_i^{k}})$ with index $k$ is the starting state of the $k$-th kinetic step, 
	and the state $({x_i^{k}}', {v_i^{k}}')$ with the index $k'$ is that of the $k$-th diffusive step, where $k=0,1, \ldots, K$. 
	The contribution of the kinetic step from $(x_i^k, v_i^k)$ to $({x_i^k}', {v_i^k}')$
	can be estimated by the so-called MC estimator, for instance, the track-length estimator \cite{luxMonteCarloParticle2018a} is used in this work. 
	In line \ref{line: mc_estimator}, $m^{\text{kinetic}}$ denotes moments contributed by the kinetic steps.
	
	\item[Diffusive part (line \ref{line: est start}--\ref{line: est end})] 
	For the diffusive simulation, in which the details of the collisions are smeared, the MC estimator, which requires the full particle information, will lead to a large bias. 
	To address this issue, we assume that the ensemble of all diffusive steps collectively constitutes a macroscopic fluid motion, governed by the fluid model \eqref{eqn: fluid-model}, since each individual diffusive step is designed to follow this model. 
	Then, we can solve the fluid model \eqref{eqn: fluid-model} and further calculate the moments \eqref{eqn: moments} )(denoted as $m^{\text{fluid}}$ in line \ref{line: fluid estimation}) as~\cite{maesHilbertExpansionBased2023a}
	\begin{align}
		\label{eqn: m0}
		m_0(x) &\approx \int_0^{\Bar{t}}\rho(x,t)dt, \\
		\label{eqn: m1}
		m_1(x) &\approx \int_0^{\Bar{t}}\rho(x,t)u_p(x) - \frac{1}{R_{cx}(x)}\partial_x\left(\sigma^2_p(x)\rho(x,t)\right)dt, \\
		\label{eqn: m2}    
		m_2(x) &\approx \int_0^{\Bar{t}}\frac{1}{2}\left(u_p^2(x)+\sigma_p^2(x)\right)\rho(x,t)- \frac{1}{R_{cx}(x)}\partial_x \left(u_p(x)\sigma_p^2(x)\rho(x,t)\right)dt.
	\end{align}
	To solve the fluid model \eqref{eqn: fluid-model}, the initial condition is given by the starting position of all diffusive steps.
	More precisely, the initial condition reads
	\begin{equation}\label{eqn: initial fluid estimator}
		\rho_f(x) = \sum_{i=1}^I \sum_{k'=0}^{K'_i} \delta(x-x_i^{k'}),
	\end{equation}
	where the delta function $\delta(x)=1$ if $x=0$; otherwise $\delta(x)=0$. 
	As for the evolution time, the durations of all diffusive steps over all particles are different. As a result, the evolution time of the fluid model, denoted as $\Theta$, is undecided.
	In this work, we approximate $\Theta$ by $\hat{\Theta}$, the average duration of all diffusive steps.
	That is, if the flight time of $i$-the particle in the $k$-th diffusive step is $\theta_i^{k}$, the empirical evolution time is then
	\begin{equation}\label{eqn: fluid estimation time}
		\hat{\Theta} = \sum_{i=1}^I \sum_{k'=0}^{K'_i} \frac{1}{I}\frac{1}{K_i'} \theta_i^{k}.
	\end{equation} 
	
\end{description}

\begin{algorithm}[h]
	\caption{KDMC simulation with the associated fluid estimation procedure}
	\label{alg: KDMC with f}
	\begin{algorithmic}[1]
		\renewcommand{\algorithmicrequire}{\textbf{Input:}}
		\renewcommand{\algorithmicensure}{\textbf{Output:}}
		\Require \parbox[t]{\linewidth}{Simulation time $\bar{t}$, time step $\Delta t$, number of particles $I$,\\
			and initial states $(x_i^0, v_i^0)$ for $i=1,\ldots,I$.}
		\Ensure Moments $m$
		
		\State $K \gets \bar{t} / \Delta t$ \Comment{Number of time steps}
		\For{$i = 1$ to $I$}
		\For{$k = 0$ to $K-1$}
		\State \label{line: kinetic start} $\tau \sim \text{Exp}\left(R(x_i^k)\right)$
		\State $\tau \gets \min(\tau, \Delta t)$ \Comment{Time of kinetic step}
		\State \label{eqn: a kinetic step} $x_i^{k'} \gets x_i^k + \tau v_i^k$ \Comment{A kinetic step}

		\State $\theta_i^k \gets \max(\Delta t - \tau, 0)$ \Comment{Time of diffusive step}
		\If{$\tau < \Delta t$} \Comment{If diffusive step exists}
		\State \label{line: kinetic end} Sample $v_i^{k'} \sim M\left(v | x_i^{k'}\right)$ \Comment{New velocity after collision}

		\State \label{line: diffusive start} Sample $\xi \sim \mathcal{N}(0, 1)$
		\State \label{eqn: a diffusive step} $x_i^{k+1} \gets x_i^{k'} + A(x_i^{k'}) \theta_i^k + \sqrt{2 D(x_i^{k'}) \theta_i^k} \xi$ \Comment{A diffusive step}
		\State Store $x_i^{k'}$ for the fluid estimation 
		\Else \Comment{If no diffusive step}
		\State $v_i^{k'} \gets v_i^{k}$ \Comment{Velocity keeps}
		\State $x_i^{k+1} \gets x_i^{k'}$ and $v_i^{k+1} \gets v_i^{k'}$ 
		\EndIf
		\State \label{line: mc_estimator} $m^{\text{kinetic}} \gets m^{\text{kinetic}} + \text{estimate}\left((x_i^k, v_i^k),(x_i^{k'}, v_i^{k'})\right)$ \Comment{using MC estimator} 
		\State \label{line: diffusive end} Store $\theta_i^k$ for the fluid estimation
		\EndFor
		\EndFor
		
		\State\label{line: est start}Construct the initial condition $\rho_f(x)$ using all $x_i^{k'}$ as Eq. \eqref{eqn: initial fluid estimator} \Comment{Fluid estimation}
		
		\State Compute the empirical evolution time $\bar{\Theta}$ using all $\theta_i^k$ as Eq. \eqref{eqn: fluid estimation time}
		
		\State \label{line: fluid estimation}\parbox[t]{\linewidth}{Solve the fluid model \eqref{eqn: fluid-model} with the initial condition $\rho_f(x)$ up to the time $\hat{\theta}$, \\
			and estimate the moments $m^{\text{fluid}}$ as \eqref{eqn: m0}-\eqref{eqn: m2}}
		\State \label{line: est end}The total moments are given by $m= m^{\text{kinetic}} + m^{\text{fluid}}$

	\end{algorithmic}
\end{algorithm}

\subsection{Overall error}\label{sec: error analysis}
In KDMC, the diffusive step (line \ref{eqn: a diffusive step} ) is derived from the fluid approximation \eqref{eqn: fluid-model}. 
This approximation is only valid when the collision rate $R(x)$ in \eqref{eqn: BBGK} is large, i.e., $R(x)=\BO(1/\varepsilon^2)$, 
where $\varepsilon\rightarrow0$ is the diffusive scaling parameter. 
Thus, the value of $\varepsilon$ determines the accuracy of the approximation of the kinetic model \eqref{eqn: BBGK} by the fluid model \eqref{eqn: fluid-model}. 
In addition, the accuracy of the KDMC simulation is also determined by the size of the time step $\dt$. 
For instance, if the time step $\dt$ is small, the KDMC simulation converges to the unbiased kinetic simulation with only a negligible statistical error. 
Therefore, the error is analyzed with respect to two limit cases: $\varepsilon\rightarrow0$ and $\dt\rightarrow0$. 

When $\varepsilon\rightarrow0$, that is, $R(x)\rightarrow\infty$, the collision rate is large and $\tau\sim\text{Exp}(R(x^0_i))$ is likely to be small. 
Within a time step $\dt$, the particle performs a kinetic step only for a short time and a diffusive step for a long time. 
In this case, $\dt\gg\tau$, which is equivalent to $\dt\gg\varepsilon^2$, the diffusive part dominates. 
Conversely, when $\dt\ll\varepsilon^2$, the kinetic part dominates. 
In conclusion, if we denote the error of the kinetic part, diffusive part, and final estimation as $\epsilon_{k}$, $\epsilon_{d}$, and $\epsilon_{kd}$, respectively, we have
\begin{equation}\label{eqn: case e_kd}
	\epsilon_{kd} = 
	\begin{cases} 
		\epsilon_k, & \text{if } \dt \ll \varepsilon^2, \\
		\epsilon_d,  & \text{if } \dt \gg \varepsilon^2.
	\end{cases}
\end{equation}
With a further analysis \cite{tang2025prep}, the error formula \eqref{eqn: case e_kd} becomes
\begin{equation}\label{eqn: err kd dt fixed}
	\epsilon_{kd}(\varepsilon) =  
	\begin{cases} 
		\BO\left(\varepsilon^2\right),  & \text{if } \dt \gg \varepsilon^2, \\
		\BO\left(\frac{1}{\varepsilon}\right),  & \text{if } \dt \ll \varepsilon^2,
	\end{cases}
	\quad \text{ and } \quad
	\epsilon_{kd}(\dt) =    
	\begin{cases} 
		\BO\left(\frac{1}{\dt}\right),  & \text{if } \dt \gg \varepsilon^2, \\
		\BO\left(\dt\right),  & \text{if } \dt \ll \varepsilon^2,
	\end{cases}
\end{equation}
where \( \epsilon_{kd}(\varepsilon) \) denotes the error of the final estimation when the time step \( \Delta t \) is fixed and the error depends only on the diffusive scaling parameter \( \varepsilon \), and \( \epsilon_{kd}(\Delta t) \) denotes that error when \( \varepsilon \) is fixed and the error depends only on \( \Delta t \).  
Next, we verify the convergence \eqref{eqn: err kd dt fixed} with a one-dimensional test case.
For simplicity, we refer to the KDMC simulation and its associated fluid estimation procedure collectively as KDMC in what follows.

\section{Numerical analysis}\label{sec: fusion test case}
In this section, we numerically verify the convergence rates given in Section \ref{sec: error analysis} and demonstrate the effectiveness of KDMC with the associated fluid estimation procedure. The analyzed method is compared with the fluid-based method from \cite{maesHilbertExpansionBased2023a}, using the kinetic MC method as a reference.

In the plasma edge region of a Tokamak, particles interact with the solid wall and undergo reflection, necessitating the use of reflective boundary conditions. The study of such boundary conditions within the KDMC framework is still an open research question, and we leave this question for future work. Therefore, we consider a test case that uses a background profile mimicking a realistic fusion scenario \cite{horstenComparisonFluidNeutral2016c}, but employs periodic boundary conditions instead.

In this test case, the particle system is simulated on the periodic domain $x\in[0, 1]$~m, from $t=0$~s to $\bar{t}=0.001$~s. The initial density is given by $\rho(x,t=0) = 1 + 1/(2\pi)\times\sin(2\pi x)$~[m$^{-3}$]. 
The variance of the plasma velocity and the collision rate are set to the values used in \cite{horstenComparisonFluidNeutral2016c}, respectively, 
\begin{equation}\label{eqn: hetero bg}
	\sigma_p^2(x) = \frac{eT_i(x)}{m_p} \quad [\mathrm{m}^2\mathrm{s}^{-2}], \quad \text{and} \quad R(x) =  \rho_i \cdot 3.2\times10^{-15}\left(\frac{T_i(x)}{0.026}\right)^{1/2} \quad [\mathrm{s}^{-1}], 
\end{equation}
where $e\approx 1.60\times10^{-19}$~J$\cdot$eV$^{-1}$ is the value of one electronvolt, $m_p\approx 1.67 \times 10^{-27}$~kg is the plasma particle mass, 
and $T_i(x) = 5.5 + 4.5\times \cos(2\pi x)$~[eV] is the temperature of ions, 
which reaches $10\,\text{eV}$ at the domain boundaries and equals $1\,\text{eV}$ at the center of the domain. The ion density $\rho_i = 10^{21}$ m$^{-3}$ is a constant. The meaning and units of the remaining constants can be found in \cite{lovbakAcceleratedSimulationBoltzmannBGK2023}.
The background \eqref{eqn: hetero bg} is modeled as being close to the limit in a diffusive scaling. To make this modeling assumption explicit, we introduce the scaled quantities
\begin{equation}
	\tilde\sigma_p^2(x) = \frac{10^{-7}}{\varepsilon^2}\sigma_p^2, \quad \text{and} \quad \tilde R(x) = \frac{10^{-7}}{\varepsilon^2} R(x),
\end{equation}
given that $\sigma_p^2(x)$ and $R(x)$ have an order of magnitude of $10^7$. In the experiment, we vary $\varepsilon$ from $10^0$ to $10^{-3.5}$, corresponding to a transition from low- to high-collisionality regimes. 
Lastly, the mean plasma velocity is chosen as $ u_p(x) = 100 + 1/(6\pi)\times\sin(6\pi x)$~[ms$^{-1}$]. 

\subsection{Convergence analysis}
\begin{figure}[h]
	\centering
	\makebox[\textwidth]{
		\begin{subfigure}{0.45\textwidth}
			\centering
			\includegraphics[trim=0 0 0 30, clip, width=\textwidth]{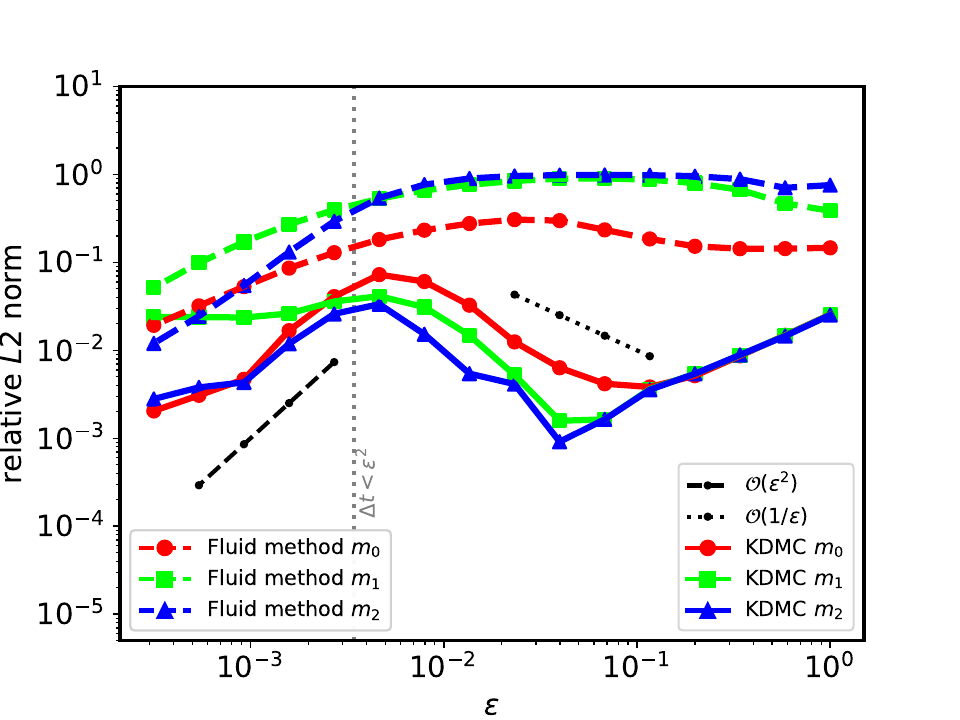}
			\caption{Fixed $\dt=\Bar{t}/85$, varying $\varepsilon$}
			\label{fig: hetero est error e}
		\end{subfigure}
		\hspace{0.02\textwidth}
		\begin{subfigure}{0.45\textwidth}
			\centering
			\includegraphics[trim=0 0 0 30, clip, width=\textwidth]{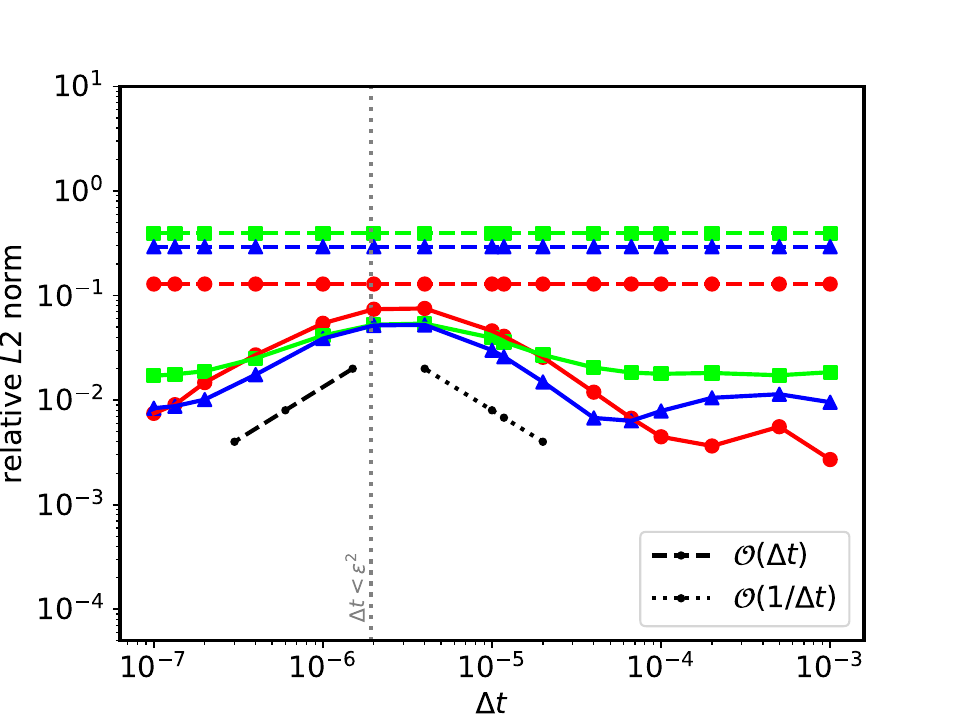}
			\caption{Fixed $\varepsilon=0.0027$, varying $\dt$}
			\label{fig: hetero est error dt}
		\end{subfigure}
	}
	\captionsetup{justification=centerlast}
	\caption{The total estimation error $\epsilon_{kd}$ of the one dimensional test case.}
	\label{fig: hetero est error}
\end{figure}

Using the kinetic MC method as a reference and the fluid-based method \cite{maesHilbertExpansionBased2023a} for comparison, the convergence is displayed in Figure~\ref{fig: hetero est error}. 
In the figure, solid lines represent the estimation errors of KDMC, while dashed lines correspond to the errors of the fluid-based method. 
The colors, red, green, and blue, denote the errors of $m_0, m_1$, and $m_2$ moments, respectively. 
The vertical dashed line, located at $\dt=\text{mean}(1/\tilde{R}(x))$, separates the two regimes: $\dt\ll\varepsilon$ and $\dt\gg\varepsilon$.

Fixing the time step $\dt=\Bar{t}/85$, the error against the diffusive scaling parameter $\varepsilon$ is shown in Figure \ref{fig: hetero est error e}.
In this test case, the error of the fluid-based method (dashed lines) starts to converge only when $\varepsilon$ is around $10^{-3}$, while KDMC has smaller errors (solid lines) for all $\varepsilon$. 
We see that the KDMC estimation error (solid lines) $\epsilon_{kd}(\varepsilon) = \BO(1/\varepsilon)$ when $\dt\ll\varepsilon^2$, and $\epsilon_{kd}(\varepsilon) = \BO(\varepsilon^2)$ when $\dt\gg\varepsilon^2$, consistent with~\eqref{eqn: err kd dt fixed}.
When $\varepsilon=10^{-3.5}$, which is the fusion test case in \cite{horstenComparisonFluidNeutral2016c}, KDMC outperforms the fluid-based method in terms of error by one order of magnitude. 
Outside the high collisional regime, KDMC is even more accurate.  
Note that the specific choice of the time step $\dt$ is made because the results are easier to interpret visually. 
The same convergence can be observed with other time step sizes.

Fixing \( \varepsilon = 0.0072 \), the error varying with \( \Delta t \) is illustrated in Figure~\ref{fig: hetero est error dt}. The analytical convergence rates from \eqref{eqn: err kd dt fixed} are clearly observable: when \( \Delta t \ll \varepsilon^2 \),  \(\epsilon_{kd}(\dt) = \mathcal{O}(\Delta t) \); whereas when \( \Delta t \gg \varepsilon^2 \), \( \epsilon_{kd}(\dt)=\mathcal{O}(1/\Delta t) \). 
For all choices of $\dt$, the error of KDMC is lower than that of the fluid method.
Especially for $\dt\geq10^{-4}$, the error of KDMC is more than one order of magnitude lower than that of the fluid method.

In conclusion, the error of KDMC follows the formula \eqref{eqn: err kd dt fixed}. The scheme outperforms the fluid-based method \cite{maesHilbertExpansionBased2023a} in terms of accuracy.

\subsection{Computational cost analysis}
To predict the computational cost of KDMC, we begin by analyzing the simulation stage when the trajectories of particles are generated. We assume that the computational cost of a kinetic step (Alg.~\ref{alg: KDMC with f}, line \ref{eqn: a kinetic step}) and a diffusive step (Alg.~\ref{alg: KDMC with f}, line \ref{eqn: a diffusive step}) is comparable, as both steps first generate one or two random numbers and update the particle's position.
Therefore, we treat each step as a unit computational operation.
The kinetic simulation updates the position along with each collision. Thus, if we assume the collision rate $R=1/\varepsilon^2$, the number of operations, denoted as $C_k$, is $C_k = R\Bar{t}=\Bar{t}/\varepsilon^2$ with $\Bar{t}$ the simulation time and $R\Bar{t}$ the total number of collisions.
KDMC has two positional updates in each $\dt$; hence, the number of operations, denoted as $C_{kd}$, is $C_{kd} = 2\Bar{t}/\dt=2K$, where $K$ is the number of time steps of size $\dt$.
The KDMC simulation has fewer operations than the kinetic simulation if $C_{kd}<C_k$, which gives the condition $\dt>2\varepsilon^2$. 
It indicates roughly the regime where the error of the diffusive part dominates. For instance, the left side of the vertical dashed line in Figure \ref{fig: hetero est error e}, and the right side of it in Figure \ref{fig: hetero est error dt}. 

We now consider the estimation stage when the moments \eqref{eqn: moments} are calculated. The moments in the kinetic MC method are estimated concurrently with the simulation, incurring negligible additional cost.
As for KDMC, the fluid estimation is performed after the simulation. This estimation solves the fluid model \eqref{eqn: fluid-model} with the evolution time $\hat{\Theta}<\dt$. In the meantime, the computational cost of solving the fluid model in the estimation stage is typically negligible compared with the cost of the KDMC simulation.
Therefore, the computational cost is decided by the simulation stage in both the kinetic MC method and KDMC. The speed-up achieved by using KDMC instead of the kinetic MC method is then effectively quantified by the ratio of operations $C_k/C_{kd}\approx \dt/\varepsilon^2$.

Table~\ref{tab: timing} presents the computation time for the test case introduced in Section~\ref{sec: fusion test case}, with $\varepsilon=0.0027$, corresponding to Figure~\ref{fig: hetero est error dt}.
The first three columns display the number of time steps, the simulation time, and the estimation time measured in seconds, respectively.  It is evident that the estimation time is negligible compared to the simulation time. 
The fourth column presents the speed-up, compared to the run-time of the reference kinetic method, which is $48066.03$ seconds. 
It can be readily verified that the speed-up is approximately proportional to $1/K$, with $K$ the number of time steps. 
As $K$ increases such that $C_{kd} > C_k$, that is, $\Delta t = \bar{t}/K \leq 2\varepsilon^2$, KDMC becomes less efficient, resulting in a slowdown indicated in the gray-highlighted rows of the table.

\begin{table}[h]
	\centering
	\caption{Runtime of KDMC simulation with fluid estimation.  Speed-up is computed by comparing the total KDMC runtime to that of the kinetic method (\(48066.03\) seconds). Slowdown cases are highlighted in gray. }
	\begin{tabular}{c|c|c|c}
		\hline
		\# time steps $K$ & Simulation time ($s$) & Estimation time ($s$) & Speed-up \\
		\hline
		1   & 1216.29  & 0.111105  & 39.47 \\
		2   & 1874.68  & 0.052020  & 25.64 \\
		5   & 3637.70  & 0.020168  & 13.22 \\
		10  & 6686.51  & 0.011620  & 7.19 \\
		15  & 9701.11  & 0.006695  & 4.95 \\
		25  & 15758.76 & 0.004488  & 3.05 \\
		50  & 30865.44 & 0.002064  & 1.56 \\
		\rowcolor{gray!20}
		85  & 51739.55  & 0.001479  & 0.93 \\
		\rowcolor{gray!20}
		100 & 60388.68  & 0.001100  & 0.80 \\
		\hline
	\end{tabular}
	
	\label{tab: timing}
\end{table}

We finish this section by remarking that the test case above is performed in a one-dimensional problem in space. 
If a two- or three-dimensional case is considered, the computational cost of solving the fluid model in the estimation stage may be significant if a large $\dt$ is used. 
Nevertheless, we expect that the cost of the KDMC simulation will still dominate.
The analysis presented in this work can guide users in choosing an appropriate time step $\dt$ for different application scenarios.

\section{Conclusion}\label{sec: conclusion}
We present a numerical error analysis for KDMC with the fluid estimation procedure, with respect to the time step $\dt$ and the diffusive scaling parameter $\varepsilon$. 
A one-dimensional test case is used to validate the given error bound that we plan to publish in the future. 
In addition, we compare the analyzed method with a fluid-based method. 
It turns out that the former has lower errors than the latter.  
The subsequent computational cost analysis shows that KDMC with the fluid estimation obtains a clear speed-up over the kinetic MC method of a factor of $\dt/\varepsilon^2$.
This work confirms the effectiveness of the estimation procedure under periodic boundary conditions. To address more realistic scenarios in\ future work, we plan to incorporate reflective boundary conditions into KDMC as well as the fluid estimation. 
	
\section*{Acknowledgments}
This work has been carried out within the framework of the EUROfusion Consortium, funded by the European Union via the Euratom Research and Training Programme (Grant Agreement No 101052200 — EUROfusion). The views and opinions expressed herein do not necessarily reflect those of the European Commission. Part of this research was funded by the Research Foundation Flanders (FWO) under grant G085922N.
Emil Løvbak was funded by the Deutsche Forschungsgemeinschaft (DFG, German Research Foundation) – Project-ID 563450842. 
The authors are also grateful to Vince Maes for valuable discussions.

\section*{Data Availability Statement}
The data and source code that support the findings of this study are publicly available at:  
\url{https://gitlab.kuleuven.be/numa/public/kdmc_and_fluid_estimation_analysis}
	
\bibliographystyle{plain}
\bibliography{your_clean}

\begin{thebibliography}{1}

\bibitem{horstenComparisonFluidNeutral2016c}
N.~Horsten, W.~Dekeyser, G.~Samaey, and M.~Baelmans.
\newblock Comparison of fluid neutral models for one-dimensional plasma edge
  modeling with a finite volume solution of the {{Boltzmann}} equation.
\newblock {\em Physics of Plasmas}, 23(1):012510, January 2016.

\bibitem{lovbakAcceleratedSimulationBoltzmannBGK2023}
Emil Lovbak and Giovanni Samaey.
\newblock Accelerated {{Simulation}} of {{Boltzmann-BGK Equations}} near the
  {{Diffusive Limit}} with {{Asymptotic-Preserving Multilevel Monte Carlo}}.
\newblock 45(4):A1862--A1889.

\bibitem{luxMonteCarloParticle2018a}
Ivan Lux and Laszlo Koblinger.
\newblock {\em Monte {{Carlo Particle Transport Methods}}: {{Neutron}} and
  {{Photon Calculations}}}.
\newblock CRC Press, 1 edition.

\bibitem{maesHilbertExpansionBased2023a}
V.~Maes, W.~Dekeyser, J.~Koellermeier, M.~Baelmans, and G.~Samaey.
\newblock Hilbert expansion based fluid models for kinetic equations describing
  neutral particles in the plasma edge of a fusion device.
\newblock 30(6):063907, 2023.

\bibitem{mortierKineticDiffusionAsymptoticPreservingMonte2022a}
Bert Mortier, Martine Baelmans, and Giovanni Samaey.
\newblock A {{Kinetic-Diffusion Asymptotic-Preserving Monte Carlo Algorithm}}
  for the {{Boltzmann-BGK Model}} in the {{Diffusive Scaling}}.
\newblock 44(2):A720--A744, 2022.

\bibitem{mortierEstimationPostprocessingStep2022}
Bert Mortier, Vince Maes, and Giovanni Samaey.
\newblock Estimation as a post-processing step for random walk approximations
  of the {{{\textsc{Boltzmann}}}}{\textsc{-}}{{{\textsc{BGK}}}} model.
\newblock 62(5--6), 2022.

\bibitem{reiterEIRENEB2EIRENECodes2005a}
D.~Reiter, M.~Baelmans, and P.~B{\"o}rner.
\newblock The {{EIRENE}} and {{B2-EIRENE Codes}}.
\newblock {\em Fusion Science and Technology}, 47(2):172--186, February 2005.

\bibitem{tang2025prep}
Zhirui Tang, Emil Løvbak, Julian Koellermeier, and Giovanni Samaey.
\newblock Analysis of kinetic-diffusion monte carlo simulation and source term
  estimation schemes in nuclear fusion applications.
\newblock Manuscript in preparation, 2025.

\end{thebibliography}
	
\end{document}